\documentclass[11pt]{article}
\usepackage{blois2002,epsfig}
\newcommand{\be}{\begin{eqnarray}}
\newcommand{\ee}{\end{eqnarray}}
\newcommand{\bi}{\bibitem}
\newcommand{\lar}{\leftarrow}
\newcommand{\rar}{\rightarrow}

\newcommand{\ds}{\partial \!  \! \! /}
\def\Journal#1#2#3#4{{#1} {\bf #2}, #3 (#4)}

\def\PRL{\em Phys. Rev. Lett.}
\def\PRD{{\em Phys. Rev.} D}

\begin{document}
\vspace*{4cm}
\title{
COSMOLOGICAL MATTER-ANTIMATTER ASYMMETRY
AND ANTIMATTER IN THE UNIVERSE
}

\author{ A.D. DOLGOV }

\address{INFN, sezione di Ferrara,
Via Paradiso, 12 - 44100 Ferrara,
Italy \\
and \\
ITEP, Bol. Cheremushkinskaya 25, Moscow 113259, Russia}

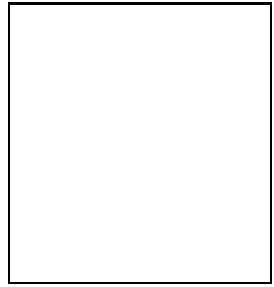
\maketitle\abstracts{
Models of baryogenesis which may lead to astronomically significant 
amount of antimatter in the universe are reviewed. Observational 
features are briefly discussed.
}

\section{Introduction}

Prediction of antimatter by Dirac~\cite{dirac28} (1928) and quick 
subsequent discovery of the first antimatter particle, positron, by 
Anderson~\cite{anderson33} (1933) are among greatest scientific 
achievements of XX century. Discovery of 
antiproton 22 years later~\cite{chamberlain55} was the next step which
had opened a whole new world of antiparticles. Now for (almost) every 
elementary particle a corresponding antiparticle has also been 
observed. The latter have opposite signs of all associated charges
and, according to CPT theorem\cite{cpt}, the same masses and, if 
unstable, life-times. Such symmetry between particles and antiparticles
naturally leads to suggestion that the universe as a whole may be also
symmetric with respect to transformation from particles to 
antiparticles and there should exist 
astronomically large regions consisting of antimatter: anti-stars,
clouds of diffuse antimatter, whole anti-galaxies, or even large
clusters of anti-galaxies. As Dirac~\cite{dirac33} 
said in his Nobel lecture~\footnote{I have found 
this quotation in the recent 
review~\cite{galaktionov02} on experimental search for cosmic 
antimatter}: ``... we must regard it rather as an accident that the
Earth (and presumably the whole solar system), contains a
preponderance of negative electrons and positive protons. It is 
quite possible that for some of the stars it is the other way about,
these stars being built up mainly of positrons and negative protons.
In fact, there may be half the stars of each kind.''  

In reality, at least in our neighborhood, the picture is different.
One sees that the nearest part of the universe is strongly matter 
dominated. Our Galaxy is definitely made of matter and only a minor
fraction of antimatter is allowed by astronomical observation. In
the case of distant galaxies, either colliding or situated in a common 
cloud of intergalactic gas, an absence of proton-antiproton 
annihilation features makes one to conclude that these objects are made
of the same form of matter (or antimatter). So the simplest conclusion
is that all the universe consists of matter only and there is
no cosmologically significant antimatter, despite CPT-symmetry
between particles and antiparticles. The mechanism that explained
the observed picture and could lead
to 100\% baryonic universe (possibly without any antimatter) was 
proposed in 1967 by Sakharov~\cite{sakharov67}.

Still there persists the question, if there may exist astronomically 
large domains or objects of antimatter? Where are they? How much
antimatter are in the universe? These problems present an interesting
and important challenge to astronomy of XXI century and to
cosmology of the early universe.

Up to the present day no traces of cosmological antimatter have been
seen. A small amount of antiprotons and positrons observed
in cosmic rays may be explained by their secondary origin in 
collisions of energetic cosmic particles or in catastrophic stellar
processes. Not a single, anti-nuclei, even as light as anti-deuterium, 
has ever been registered. According to the analysis of the data
on gamma-ray background with energy around 100 MeV, made in the
review~\cite{steigman76}, the nearest anti-galaxy should be away
at least at 10 Mpc:
\be
l_B > 10\,\,{\rm Mpc}
\label{lB1}
\ee
A much stronger limit was obtained relatively recently by Cohen,
De Rujula, and Glashow~\cite{cohen98} who concluded that
the nearest part of the universe dominated by antimatter should be
at least at Gigaparsec distance scale. The result was obtained
for baryo-symmetric universe (equal amount of matter and antimatter)
and for the so called adiabatic density perturbations. Lifting these
constraints may allow to weaken this very restrictive limit (see 
discussion below). 

Despite these bounds, an existence of astronomically large objects of 
antimatter (gas clouds, anti-stars, antigalaxies,...) is not excluded
and even theoretically natural. They may be as close as halo of the
Galaxy and rich of heavy anti-nuclei, e.g. $\bar C$, $\bar N$,
$\bar O$, ... and maybe $\bar Fe$. It is evidently very interesting to
look for such anti-objects. Their discovery would lead to better
understanding of the nature of C and CP violation in cosmology, 
would allow to get an insight into mechanisms of baryonic charge 
non-conservation, and may give another proof (or evidence) of
inflation.

\section{The problems to think about}

The excess of baryons over antibaryons is usually presented as
the dimensionless ratio:
\be
\beta = {N_B - N_{\bar B} \over N_\gamma} \approx 6\cdot 10^{-10}
\label{beta}
\ee
where $N_{B,\bar B,\gamma}$ are the cosmological number densities
of baryons, antibaryons, and photons in cosmic microwave background
radiation (CMBR). The last number is precisely determined from direct
observations, $N_\gamma = 410/{\rm cm}^3$, while the baryon number
density is independently found from big bang nucleosynthesis 
(see e.g.~\cite{dolgov02}) 
and from relative positions and heights of acoustic peaks in angular
spectrum of CMBR (see e.g. ~\cite{cmbr-par}).

At the present day $N_{\bar B} \ll N_B$, but in the early universe 
when the temperature was above QCD phase transition, $T>100$ MeV,
number of baryons was very close to the number of the antibaryons,
$N_{\bar B} =  N_B(1-\beta)$. The quantity $\beta$ is often called
baryon asymmetry of the universe.

There are a few ``five-star'' questions to be addressed in connection
with baryon asymmetry:
\begin{enumerate}
\item{}
Is $\beta = const$ or it may change in different space points, i.e.
baryon asymmetry may be inhomogeneous?
\item{}
Is $\beta >0$ and the universe is dominated by matter everywhere
or somewhere $\beta <0$ and there are regions of antimatter?
\item{}
Is total baryonic charge of the universe:
\be
B_{tot} = \int \beta\, dV 
\label{B-tot}
\ee
positive, negative, or exactly zero? The last case is particularly
appealing because it means that the universe is globally charge
symmetric, while locally and at a very large scale it is strongly
asymmetric.
\item{}
If baryon asymmetry is not homogeneous, i.e. $\beta = \beta (x)$,
then what is the characteristic scale of its variation, $l_B$?
\item{}
What kind of anti-objects may exist?
\end{enumerate}

Data possible exclude $B_{tot}=0$ inside horizon, 
$l_{hor} \sim 3$ Gpc,
as was mentioned above, but all other questions remain unanswered.

\section{Principles of baryogenesis}

The idea of dynamical generation of baryon asymmetry of the
universe based on possible non-conservation of baryon number
was pioneered by Sakharov~\cite{sakharov67}. Two years later 
another suggestion was put forward by Omn{\'e}s~\cite{omnes69}
that the observed cosmological baryon asymmetry is only local
and that it was created by spatial separation of baryons and 
antibaryons on astronomically large scales. It can be realized
if baryons are conserved. However it seems
impossible to create such separation and the Omn{\'e}s idea is 
abandoned now. Maybe it may revive in a modified form if one
finds a mechanism for spatial separation of baryons and antibaryons
along orthogonal direction to our 3-dimensional space, into higher
dimensions, but at the moment it looks too far-fetched.

According to Sakharov three basic principles of baryogenesis are 
the following:
\begin{enumerate}
\item{}
{\bf Non-conservation of baryons.}\\
It was the weakest point in 1967 when the common belief was that
baryons are conserved: ``we exist, hence protons must be stable''.
Since that time our understanding became opposite: 
``we exist, hence protons must be unstable'', because most
probably the universe suitable for life could not be created if
baryonic charge is conserved. Present day theory has a favorable
attitude to non-conservation of baryons: both electroweak
theory and grand unification ones predict that baryonic charge is
not conserved.
\item{}
{\bf Violation of charge symmetry, both C and CP.}\\
Both C and CP breakings are well established in experiment and
though the concrete mechanism of CP-breaking is unknown it can be
easily implemented into theory by introducing complex mass matrices
of particles or complex Yukawa coupling constants. 
\item{}
{\bf Deviation from thermal equilibrium.}\\
In thermal equilibrium occupation number in phase space is given
by Bose or Fermi (depending on particle spin) distribution
functions:
\be
f = \left( 1 \pm e^{E/T}\right)^{-1}
\label{feq}
\ee
where $E$ is the particle energy and $T$  is the temperature. 
Chemical potentials must vanish in equilibrium if the corresponding
charge is not conserved. Due to CPT-theorem the masses of particles
and antiparticles are equal and thus for equal values of
particle momenta their energies must be equal too. Correspondingly
$f=\bar f$ and in thermal equilibrium numbers of particles and 
antiparticles must be the same. If CPT is broken then charge
asymmetry may appear even in thermal equilibrium~\cite{dolgov81}.

However breaking of CPT is not necessary because the universe is
non-stationary, it expands and for massive particles some deviations
from equilibrium are always present. Also a
noticeable breaking of thermal equilibrium could arise as a result
of possible first
order phase transitions in the primeval cosmic plasma in the
process of expansion and cooling down.
\end{enumerate}

Thus all necessary conditions for baryogenesis are naturally 
fulfilled and one may expect that some asymmetry between particles
and antiparticles should be created in the universe. Moreover,
none of three formulated above conditions are obligatory and in
some (possibly rather exotic) models cosmological baryon asymmetry 
might be generated even in thermal equilibrium, or with
conserved baryonic charge in particle physics, or in CP-invariant
particle theory - examples of such scenarios are given below.

\section{Mechanisms of CP-breaking in cosmology \label{s-cp}}

There are three possible ways of breaking symmetry between particles 
and antiparticles in cosmology. One is a standard 
{\it explicit} mechanism
when  CP-breaking is achieved by complex constants in the Lagrangian.
In this case baryon asymmetry in the universe would be directly 
related to CP-violating processes in elementary particle physics.
Baryon asymmetry in this case would be the same homogeneous
one over all the universe, $\beta = const$.

Another mechanism is {\it spontaneous} CP-breaking, proposed by
T.D. Lee~\cite{lee73}. It can be achieved by vacuum condensate of 
a complex scalar field, $\langle \phi \rangle$,
which introduces complex constants into
Lagrangian over the vacuum state with a certain non-zero value of 
$\langle \phi \rangle$. For example, the Yukawa interaction term
$ g\phi \bar\psi \psi$ gives rise to the complex mass term
$m_\psi = g \langle \phi \rangle$. One needs, of course, several
different terms with $\phi$ in the Lagrangian so that the phase
of $\langle \phi \rangle$ could not be rotated away by redefinition
of the fields.

Also in this case CP-violation observed in particle physics can be
related to the cosmological charge asymmetry but now $\beta$
cannot be spatially constant. Spontaneous breaking of a discrete
symmetry (as e.g. CP) would give rise to domain structure of the
universe with a different signs of $\langle \phi \rangle$ in 
different domains. This might lead to cosmological 
disaster, as was shown by Kobzarev, Okun and 
Zeldovich~\cite{zeldovich74}. The problem is that domains with
different sings of CP-breaking amplitude $\langle \phi \rangle$
must be separated by domain walls with a huge surface energy 
density and even a single domain wall per horizon would destroy
the observed isotropy of the universe. If there are too much walls
per horizon the universe would be over-closed. To solve the domain 
wall problems one should either invent a mechanism enlarging the 
size of the domains beyond the present day cosmological horizon or 
to destroy the walls at an early stage of cosmological evolution. 

Evidently in the domains with different signs of 
$\langle \phi \rangle$ the sign of CP-violation would be different
and correspondingly either an excess of matter over antimatter would 
be created or, another way around, antimatter over 
matter~\cite{zeldovich74,brown79}. 
With such mechanism of CP-breaking the 
universe would be globally charge symmetric, $B_{tot}=0$. However
in a simple version of the model the size of the domains should be
much smaller than galactic size in evident contradiction with the 
data. A way out was suggested by K. Sato~\cite{sato81} who assumed
that there could be a moderate exponential expansion to inflate
the domains to astronomically interesting scales. However,
the problem of domain wall still has to be solved. Even if the 
walls are eliminated by one or other mechanism, matter and 
antimatter remain in close contact and according to the arguments
of ref.~\cite{cohen98} the nearest antimatter region should be
practically at the horizon distance, around Gpc.

There could be another mechanism of CP-violation which operated only
in the early universe and is absent in the particle physics. This
mechanism may be called {\it stochastic} breaking of charge 
symmetry~\cite{dolgov92}. In a sense it is similar to spontaneous
one but does not suffer from the domain wall problems and not
necessary leads to globally charge symmetric universe. Such mechanism
could be realized if a complex scalar field did not relax down
to its equilibrium value at some stage of cosmological evolution. It
maybe be a result of a phase transition or of infrared instability
of light scalars in De Sitter (inflationary) stage~\cite{infra}. If
such a field remains non-zero during baryogenesis its amplitude would
give rise to CP-breaking in particle interactions, while later  it 
would relax down to zero and would not contribute to CP-breaking. 
Since we know that CP is broken in particle physics, then both
stochastic and explicit mechanisms might operate at baryogenesis.

\section{Models of Baryogenesis \label{s-bg}}

I will briefly describe below possible models of baryogenesis that
exist in the literature. More detailed discussion can be found e.g.
in the review papers~\cite{dolgov92,bs-rev}.

\subsection{GUT-baryogenesis or heavy particle decays \label{ss-gut}}

This is historically first model of baryogenesis, it was practically
proposed in the pioneering paper~\cite{sakharov67} and renewed
interest to it was stimulated by the paper~\cite{yoshimura78} 
a decade later. According to Grand Unified Theories 
(GUT)~\cite{georgi74} there
exist superheavy gauge of Higgs bosons whose decay do not conserve 
baryonic charge. For example such bosons can decay into the channels
$X\rar qq$ or $X\rar \bar q l$, where $q$ is a quark and $l$ is a 
lepton. Due to possible CP-nonconservation the decay branchings
$B(X\rar qq)$ and $B(\bar X\rar \bar q \bar q)$ could be different
and in out-of-equilibrium situation such decays would produce an
excess of baryons over antibaryons (or vice versa - the sign of CP
breaking at GUT scale is unknown).

The deviation from equilibrium is determined by the ratio of the
universe expansion rate, when the temperature was close to X-bosom 
mass, $m_X$, to the decay width of X-bosons:
\be
\epsilon = {H(T=m_X) \over \Gamma_X }
\approx {m_X \over \alpha\, m_{Pl}}
\approx {m_X \over 10^{17}\,{\rm GeV}}
\label{epsilon}
\ee
where $\alpha \sim 10^{-2}$ is the gauge coupling constant.
Thus, if X-bosons are as heavy as $10^{16}$ GeV, as we believe now,
the deviation from equilibrium could be quite large to produce
significant baryon asymmetry even with entropy dilution by 
approximately two orders of magnitude in the course of cosmological
cooling down from $10^{16}$ GeV to the present state.

In simple versions such mechanism naturally gives $\beta = const$ 
and no antimatter.
A problem with its realization is that probably the universe was
never at very high temperatures about $10^{16}$ GeV, since
normally after
inflation the universe was heated to much lower temperature.

\subsection{Electro-weak baryogenesis \label{ss-ew}}

It was suggested by Kuz'min, Rubakov, and Shaposhnikov~\cite{kuzmin85}
that baryon asymmetry might be generated at much lower temperatures,
$\sim$TeV, by the usual electro-weak interactions. Indeed, all
the necessary ingredients are present in the standard low energy 
physics. It is known that CP is broken, baryonic charge is 
non-conserved due to quantum chiral anomaly~\cite{thooft76}, and
thermal equilibrium would be broken if the electro-weak phase 
transition is first order. It is evident from eq. (\ref{epsilon})
that deviation from equilibrium due to non-zero masses of $W$ and
$Z$ bosons is negligibly small. Generation of baryon asymmetry
in this model takes place not in massive particle decays but
on the boundary between two phases with unbroken and broken
$SU(2)\times U(1)$-symmetry of weak interactions respectively. 
In the high temperature phase, where the symmetry is unbroken, 
sphaleron~\cite{sphaleron}
transitions, which break baryon conservation, are believed to be
very active, so equilibrium with respect to B-nonconserving
processes is quickly established. To be more exact, anomalous
electroweak interactions conserve the difference between leptonic
and baryonic charge, $(B-L)$, while $(B+L)$ is non-conserved. So
$(B-L)$ remains constant in comoving volume and equal to its 
initial value, while non-conserved $(B+L)$ may evolve. 
In low temperature phase the baryonic charge is practically 
conserved but if the phase transition was first order and phases
coexisted, in the boundary between them both baryon non-conservation
and deviation from equilibrium could be strong enough to generate the 
observed asymmetry. 

However, as we now know from the LEP data (see e.g.~\cite{pdg}), 
the lower limit on the Higgs boson mass is higher than 100 GeV
and this makes first order phase transition quite problematic.
There exist many modifications of the minimal scenario of 
electroweak baryogenesis but the model became much less popular
now. For review one can see refs.~\cite{bs-rev,ew-rev}; a 
discussion of the latest development and an impressive list of 
references can be found in the recent papers~\cite{carena02}.

Normally electroweak models do not lead to creation of
antimatter domains in the universe but with a simple modification
they could do that.

\subsection{Baryo-thru-lepto-genesis \label{ss-bl}}

This mechanism presents a combination of the two discussed above 
(subsections~\ref{ss-gut} and \ref{ss-ew})
and was proposed by Fukugita and Yanagida~\cite{fukugita86}.
Initially lepton asymmetry, $L_{in}$, was generated by decays of 
heavy Majorana neutrino, $\nu_M$, and later this asymmetry was 
redistributed between baryonic and leptonic ones
by electro-weak interactions at equilibrium stage due to
non-conservation of $(B+L)$. Final baryon asymmetry would be
roughly equal to $L_{in}/2$ (the exact value of the coefficient
depends upon the particle content of the 
theory~\cite{aoki86,dolgov92}). This scenario is very popular now,
one can find detailed discussion of different versions and a list 
of references in the reviews~\cite{lep-rev}.

Lepton asymmetry generated by the decays of $\nu_M$ can be 
roughly estimated as:
\be
L=\epsilon_M\,{N_{\nu_M} \over N_{tot}}\,{\Delta \Gamma \over \Gamma}
\label{L}
\ee
where $\epsilon_M = m_{\nu_M}/(\alpha_M m_{Pl}$ is the parameter
characterizing deviation from thermal equilibrium (\ref{epsilon}),
$N_a$ are number densities of the corresponding particles,
$\Gamma\sim \alpha_M m_{\nu_M} $ 
is the total decay width of $\nu_M$, and 
$\Delta \Gamma \sim \alpha_M^2 m_{\nu_M}\,\delta$
is the difference of lepton non-conserving decay rates of $\nu_M$
into charge conjugated channels with $\delta$ being the amplitude
of CP-breaking. With $m_{\nu_M} \sim 10^{10}$ GeV the lepton asymmetry
seems to be too small but experts in the field are more optimistic
and according to majority of the works on the subject, the discussed
mechanism may successfully generate observed baryon asymmetry of the 
universe.

Recently a modification of ``lepto-thru-baryo'' scenario was 
proposed in ref.~\cite{fukugita02}. It was suggested that 
$(B+L)$-asymmetry was generated at GUT scale by heavy particle
decays, while $(B-L)$ remained zero since it is conserved in simple
models of grand unification. Lepton asymmetry was washed out in 
equilibrium reactions
with Majorana neutrino at a later stage and remaining baryon 
asymmetry cannot be completely destroyed by sphaleron processes
since they conserve $(B-L)$.

All these models are not particularly good for creation of cosmic
antimatter but one can easily modify them to this end by introducing
e.g. stochastic CP-violation (see sec.~\ref{s-cp}). Still it looks
rather unnatural.

\subsection{Black hole evaporation \label{ss-bh}} 

Cosmological baryon asymmetry might be created by the 
evaporation of primordial black holes with low 
mass~\cite{hawking75,carr76,zeldovich76}. A concrete mechanism
was suggested in the paper~\cite{zeldovich76} and the calculations 
of the effect were performed in ref.~\cite{dolgov80}. This 
mechanism does not demand direct non-conservation of baryonic 
charge in particle physics. On the other hand, according to
estimates made by Zeldovich~\cite{zeldovich76} proton may
decay through formation of virtual black hole with life-time
about $m_{Pl}^4 / m_p^5 \sim 10^{52}$ sec. If this estimate
is correct then in the high dimensional models with TeV-scale
gravity~\cite{dvali98} the proton life-time would be 
catastrophically short, $\sim 10^{-8}$ sec. A lower bound on
quantum gravity scale based on the analysis of proton decay
through formation of virtual black holes has been recently
done in the paper~\cite{adams01}.

In the process of evaporation black hole emits all kind of 
particles with the mass smaller than black hole temperature, 
$T_{BH}= m_{Pl}^2 /(8\pi M_{BH})$.
A massive meson, still in gravitational field of
black hole, could decay into a light baryon and a heavy antibaryon
(e.g. $u$ and $\bar t$ quarks) or vice versa. The decay 
probabilities may be different because of C(CP) violation. Since 
back capture of heavy particles by the black hole is more probable 
then that of
light ones, such process could create a net flux of baryonic 
charge into external space, while equal anti-baryonic charge
would be hidden inside disappearing black hole. This mechanism 
could explain the observed value of the baryon asymmetry of the
universe if at some early stage the total cosmological energy density
was dominated by those black holes~\cite{dolgov80,dolgov00}.

Without special efforts this model is also inefficient for creation
of cosmologically significant amount of antimatter.

\subsection{Spontaneous baryogenesis \label{ss-spnt}}

The mechanism was proposed by Cohen and Kaplan~\cite{cohen87} 
and is based on the assumption that there exists a 
$U(1)$-symmetry, related to baryonic or some other
non-orthogonal charge, which is spontaneously broken. A toy-model
Lagrangian can be written as:
\be
{\cal L} = - |\partial \phi |^2 +
\lambda \left( |\phi|^2 - f^2 \right) + 
\sum_a \bar \psi_a \left(i\ds + m_a  \right) \psi +
\sum_{a,b} \left(g_{ab} \phi \bar \psi_b  \psi_a + h.c. \right)
\label{unbrkn}
\ee
where some fermionic fields $\psi_b$ possess non-zero baryonic charge,
while some other do not. The theory is invariant with respect to 
simultaneous phase rotation: 
\be 
\phi \lar \phi \exp (i\theta)\,\,\, {\rm and}\,\,\,
\psi_b \rar \psi_b \exp (i\theta) 
\label{rot}
\ee
which ensures conservation of ``baryonic'' charge. In the broken
symmetry phase where $|\phi| = f$ the conservation of baryonic
current of fermions also breaks down (due to presence of the last
term in the Lagrangian above) and the Lagrangian takes the form:
\be
{\cal L} = - f^2 \left( \partial \theta \right)^2 + 
\partial_\mu \theta\, \bar \psi_b \gamma_\mu \psi_b + ...
\label{broken}
\ee
where $\theta$ is the massless Goldstone field induced by the breaking
of the global $U_b$-symmetry. If there are some additional terms
in the Lagrangian producing an explicit symmetry breaking then
$\theta$ would be massive and is called pseudo-Goldstone field. 
Baryogenesis would be much more efficient in the latter case. 

In the homogeneous case when $\theta = \theta (t)$ the second term
in expression~(\ref{broken}) looks like $\dot \theta n_b$ where
$n_b$ is the baryonic charge density. So it is tempting to identify
$\dot \theta$ with chemical potential of baryons~\cite{cohen87}.
Though it is not exactly so~\cite{dolgov95}, still this term shifts
equality between number densities of quarks and anti-quarks even 
in thermal equilibrium, allowing baryogenesis without deviation
from thermal equilibrium. In a sense this mechanism is equivalent to
breaking of CPT because the process goes in time dependent background.

The sign of the created baryon asymmetry is determined by the
sign of the $\dot\theta$ and could be both positive or negative
producing baryons or antibaryons. To create the matter/antimatter
domain of astronomically large size the $U(1)$-symmetry should be
broken during inflation and in this case the sign of $\dot\theta$
would be determined by the chaotic quantum fluctuations at
inflationary stage. The analysis of density perturbations created
by the fluctuating field $\theta$ was done in ref.~\cite{turner89}.

In this scenario C(CP)-violation is not necessary
for generation of baryon asymmetry. As a whole the universe could be
charge symmetric. We know however that an explicit C(CP)-violation
is also present. If it also participates in creation of baryon
asymmetry, then the amount of matter and antimatter in the universe 
would be unequal with unknown ratio that should be determined from 
observations.

\subsection{SUSY condensate baryogenesis \label{ss-susy}}

All known baryons are fermions but if there is supersymmetry then
scalar superpartners of baryons should exist. Such scalar fields
might form a condensate in the early universe and accumulate a large
baryonic charge. A very efficient scenario of baryogenesis based on
this observation was proposed by Affleck and Dine~\cite{affleck85}.
In many supersymmetric models self-potential of scalar
fields has so called flat directions along which it does not rise. As
a toy model we can consider the potential of the form:
\be
U(\chi) = \lambda \left[ |\chi|^4 - \left( \chi^4 + \chi^{*4}
\right)/2 \right] = \lambda |\chi|^4 \left( 1 - 4\cos \theta
\right) 
\label{uofchi}
\ee  
where $\chi = |\chi| \exp (i\theta)$. This potential has four 
valleys $\theta = \pi n/2$ with $n=0,1,2,3$ and is not
invariant with respect to rotation in two-dimensional
$({\rm Re}\chi,\,\,{\rm Im}\chi)$-plane. This leads to
non-conservation of baryonic current of $\chi$. The latter is
defined as
\be
J_\mu^B (\chi) =(-i/2)\left( \chi^* \partial_\mu \chi -
\partial_\mu \chi^* \chi \right) = \partial_\mu \theta\,|\chi|^2 
\label{JchiB}
\ee
Initially the field $\chi$ could be far away from the origin along
one of the valleys, either by unspecified initial conditions or
due to rising quantum fluctuations at inflationary 
stage~\cite{infra}, which lead to 
$\langle \chi^2 \rangle = H^3t /2\pi$. When inflation is over and 
due to some symmetry breaking, $\chi$ acquires mass, it starts to
move down to the origin with baryonic charge accumulated in its
angular motion, $B_\chi = \dot \theta\, |\chi|^4$. 
There is a convenient mechanical analogy in the case of homogeneous
$\chi = \chi (t)$. Its equation of motion
\be
\ddot \chi + 3 H\dot \chi + U' (\chi) = 0
\label{ddotchi}
\ee
describes classical mechanical motion of a point-like particle in
the potential $U(\chi)$. The second term, induced by the cosmological
expansion, presents liquid friction force. In this language the
baryonic charge of $\chi$ is simply angular momentum of the particle
in this potential. If the potential is spherically symmetric (in two
dimensions: $[Re\chi,Im\chi]$ ) then baryonic charge 
is conserved, otherwise not. 

When $\chi$ approaches zero it decays into quarks and transfers 
baryonic charge of its angular motion to baryonic charge of 
quarks. The sign of the asymmetry depends upon the direction of
the rotation. The latter may be determined by chaotic initial
fluctuations in the directions orthogonal the the valley, in 
this case CP-violation is unnecessary, or by explicit CP-breaking
terms in the potential, e.g.
\be
U(\chi) = \lambda |\chi|^4 \left( 1 - \cos 4\theta \right)
+m^2 |\chi|^2 \left[ 1 - \cos \left( 2\theta + 2\gamma \right)\right]
\label{uodd}
\ee
where a non-zero phase $\gamma$ induces explicit CP-violation. 

If $\chi$ condensed along the valley with $\theta =0$ it would
start rotating clock-wise when it approaches zero and starts to 
feel mass valley (for $0<\gamma<\pi/2$). If $\chi$ ``lived'' in
$\pi/2$-valley it would start rotating anti-clock-wise. Thus,
depending upon initial position of $\chi$,
both baryons and antibaryons may be created with the same sign
of CP-violating phase $\gamma$. In this model the universe most
probably would be globally charge asymmetric because the 
magnitude of the angular momentum depends upon the relative angles 
between $\chi^4$ and $\chi^2$ valleys. The size of baryonic and
anti-baryonic domains are expected to be very large, maybe even
larger than horizon, because the
regions of initial values of $\chi$ in different valleys could be
strongly separated during inflation.

On the other hand, if all the observed part of the universe 
originated from the same $\chi$ valley the size of regions with
definite sign of baryon asymmetry could be much smaller. Such
situation could be realized in the model with potential (\ref{uodd})
with $\gamma =0$, i.e. with conserved CP. If $\chi$ evolves down
along the valley with $\theta = \pm \pi/2$, then for sufficiently
small $|\chi|$ this direction becomes unstable and $\chi$ would go
down to $\theta =0$ or $\theta = \pi$ rotating either clock-wise
or anti-clock-wise. So in this model there would be equally
probable baryonic and anti-baryonic domains. Slightly changing
direction of mass valley, taking a non-zero $\gamma$ would allow
to create the universe with arbitrary fraction of baryons with 
respect to antibaryons.

\section{Creation of domains or objects of antimatter}

As we have seen above, antimatter could be quite naturally created
in the universe in astronomically interesting amount. Probably the
most efficient mechanisms for creation of cosmic antimatter are
spontaneous baryogenesis (sec.~\ref{ss-spnt}) and Affleck-Dine 
one (sec.~\ref{ss-susy}). If the scalar baryon of SUSY model
is coupled to the inflaton field $\Phi$ by the following
general renormalizable coupling
\be
{\cal L} =  \left( \lambda \chi^2 + h.c.\right)
\left( \Phi - \Phi_1 \right)^2
\label{chi-phi}
\ee
a very interesting and unusual features in generation of baryon 
asymmetry may appear~\cite{dolgov93}. Because of this 
interaction favorable conditions for baryogenesis might be created
only for a short time and correspondingly in relatively small
fraction of space. As a result there would be separate bubbles
with very large baryon asymmetry, even of order unity which would
be floating in the usual cosmological baryo-asymmetric background 
with $\beta = 6\times 10^{-10}$. This baryon-rich bubbles would be
with practically equal probability
baryonic or anti-baryonic and many of them
would form primordial black holes and because of that would be
indistinguishable. Some of them would survive collapse and form
disperse clouds of matter and antimatter with unusually high
baryonic number. Some of this objects may form unusual stars and 
anti-stars with a high initial fraction of heavy nuclei (or anti-nuclei).
The mass distribution  of (anti)baryon-rich regions,
according to ref.~\cite{dolgov93}, is
\be
{dN \over dM} \sim \exp \left( -C\,\ln^2 {M\over M_1}\right)
\label{dndm}
\ee
where $M_1$ is unknown mass parameter which could be a fraction 
of solar mass up to many solar masses and $C$ is a constant.

It is an interesting possibility that such primordially formed black 
holes constitute dark matter in the universe. It would be a cold
dark matter but, in contrast to the standard one, having a disperse
mass distribution. With a small probability there could be even very
heavy black holes with the masses up to $10^6-10^9\, M_\odot$,
where $M_\odot$ is the solar mass. Most of these black holes might
have mass near solar mass and could live in galactic halos. Their
non-captured atmospheres may be anti-baryonic and, if so, considerable
amount of antimatter could be even in halo of the Galaxy. The model
naturally predicts early formation of quasars and evolved chemistry
in their atmospheres because in the (anti)baryon-rich regions
primordial nucleosynthesis would not stop at $^7Li$. Regions with
a smaller $\beta$ might survive collapse and form either clouds of
antimatter or anti-stars. Encounter of a star and anti-star could be 
a very spectacular phenomenon. More details about this model
can be found in refs.~\cite{dolgov01}.
 
Still a few comments about formation and properties of such 
antimatter black holes may worth making. They were formed when
the density contrast inside horizon was of order unity,
$(\delta \rho /\rho)_{hor} \sim 1$. One can simply estimate
this ratio as
\be
{\delta \rho \over \rho } \approx {m_p \over 3T}\,
{N_B \over N_\gamma} = {1\over 3}\, \beta \, {m_p \over T}
\label{delta-rho}
\ee
where $m_p$ is the proton mass and $T$ is the cosmological
temperature. The mass inside horizon is given by:
\be
M_{hor} = {8\pi \over 3}\, \left( 2t\right)^3 \rho =
5\cdot 10^4 M_\odot \, \left( {t \over {\rm sec}}\right)
\label{m-hor}
\ee

Black holes formed in antibaryon rich regions could possess atmosphere 
of antimatter. Before hydrogen recombination diffusion is slow
and the atmosphere might survive. Recombination in high $|\beta|$ 
regions occurs earlier, at temperatures in eV range, instead of
normal 0.1 eV. Hence atmosphere would not be stripped by CMBR at
the onset of structure formation. Annihilation on the boundary
could heat the plasma and create secondary ionization. All these
phenomena deserve more detailed investigation.

An efficient way to create cosmic antimatter in any model of
baryogenesis is based on stochastic CP-breaking~(sec.~\ref{s-cp}).
If some complex scalar field was nonvanishing during baryogenesis
then effective CP-breaking induced by this field would participate
in creation of cosmological charge asymmetry and if the field
changed sign as a function of space point the sign of asymmetry might
also be different. For example, if the field $\phi (t,x)$ was
a slowly varying function of $x$ and its potential was not strictly
harmonic, i.e. contained e.g. quartic terms, $\lambda \phi^4$, and
since frequency of oscillations in non-harmonic potential depends
upon the amplitude, then smooth $\phi(x,t)$ would turn into an
oscillating (and quasi-periodic) function of space point. Depending  
upon the ratio between amplitudes of explicit and stochastic 
CP-violation astronomically interesting antimatter domains might be 
formed~\cite{dolgov86}.

Discussion of some other models of antimatter creation, can be found 
in the papers~\cite{dolgov92,dolgov93a}. This subject is also actively
discussed in this conference~\cite{anti-blois}.

\section{Bounds on existence of antimatter }

Let us first discuss the most stringent lower bound on the distance
to antimatter domains in baryo-symmetric cosmology
found by Cohen, De Rujula, and Glashow~\cite{cohen98}. Matter and
antimatter cannot be separated by the distance, $l_B$, larger than 
20 Mpc (at the
present day scale) because in the boundary region with deficit of
both matter and antimatter density contrast would be incompatible
with measurements of angular fluctuation of CMBR accessible at such
scales. If $l_B < 20$ Mpc diffusion of protons and antiprotons would
bring them into contact and they would annihilate. Naively one would
expect that the excessive pressure created by annihilation would
force matter and antimatter apart inhibiting the annihilation. 
However the picture is opposite. The energetic products of annihilation
have a large mean free path so they transfer their energy (and 
pressure) to the plasma far away and the excessive pressure pushes
matter and antimatter together and enhances annihilation. Thus the
CDRG~\cite{cohen98} limit $l_B > (\sim {\rm Gpc})$ was obtained.
As it was already mentioned it is valid in the case of baryo-symmetric
cosmology when the universe is equally rich of baryons and antibaryons.
If however, the fraction of antibaryons is much smaller than that of
baryons (or vice versa) the limit would be weaker, roughly by the
same amount. 

Another assumption made in the derivation of CDRG bound is
adiabatic density perturbations, i.e. the same perturbations in
all forms of matter. The limit may be invalid for isocurvature 
perturbations, when initially $\delta \rho =0$ but chemical content
of plasma varies in different space points. Now let us consider 
three regions: one with baryonic excess, another with equal 
anti-baryonic excess and some boundary between them with no baryons.
By assumption the energy density of all three regions were initially
the same. The plasma in baryonic and anti-baryonic regions would be
less relativistic and thus, when the temperature drops, their energy 
density would be higher than the energy density of the baryon-free 
part. On the other hand, nonrelativistic matter cools faster and the
temperature of radiation in baryon free region would be higher. Hence
the pressure of photons would be higher and it may prevent baryons
and antibaryons from close contact and annihilation. This mechanism
may significantly relax CDRG bound, but accurate calculations are
needed. 

Matter-antimatter annihilation on the domain boundaries might distort
energy spectrum of CMBR if the annihilation takes place before
recombination~\cite{kinney97}.
The effect would be noticeable for large size domains. 
At the present time
this phenomenon does not permit to obtain any interesting limit. 

Of course an unambiguous proof of existence of cosmic antimatter would
be an observation of anti-nuclei in cosmic rays, especially heavier
than anti-deuterium because the latter might be produced in secondary 
processes~\cite{donato99} with small but not vanishingly small
probability. Some indication for antimatter could be anomalies in
the spectrum of antiprotons in cosmic rays but the data may have an
ambiguous interpretation. Earlier works on search of cosmic antiprotons
and their relation to cosmological antimatter can be found 
in refs.~\cite{chechetkin82z}.
Program of search and existing limits
are discussed in many talks at this Conference, too many to mention
them all, but for reviews one can see 
refs.~\cite{galaktionov02,anti-rev}.

\section{Conclusion}

Simplest models of baryogenesis predict charge asymmetric universe
with $B_{tot} \neq 0$. Normally, without additional assumptions, they
do not lead to cosmological antimatter domains or 
astronomically large objects. On the other 
hand, they generally meet some problems with explanation of the 
observed value of baryon asymmetry, $\beta$. 

Spontaneous baryogenesis or Affleck-Dine scenario can successfully
explain the observed asymmetry and naturally predict existence of
astronomically interesting antimatter.

In a version of Affleck-Dine model with coupling between the scalar 
baryon and inflaton fields relatively small regions with very high
$\beta$, both positive and negative, can be created. Such objects 
would mostly form primordial black holes. They may be abundantly present 
even in halo of our Galaxy. An interesting possibility is that they 
form all dark matter in the universe. Around such black holes heavy
(anti)nuclei should be present and, though one cannot distinguish
between black hole and anti-black-hole, anti-nuclei stripped from
the neighborhood of the latter may be registered. Thus one may hope to
discover anti-nuclei in cosmic rays and not only $^4 \bar He$ but
also (and quite probably more abundant) heavier anti-nuclei, $\bar C$,
$\bar N$, $\bar O$ and maybe even $\bar Fe$. 

Other models give less optimistic scenarios for possibility of 
registration of anti-nuclei. But there are no iron-strong predictions
and one should rely on future observations and detailed
theoretical calculations to make more definite conclusions. 
At the present day our knowledge is too scarce and one even
cannot exclude that we live in antimatter dominated universe
with minor domains of matter as e.g. our group of galaxies inside
10 Mpc.

\end{document}